\documentclass[paper]{ieice}
\usepackage[pdftex]{graphicx,xcolor}% for pdflatex
\usepackage[fleqn]{amsmath}
\usepackage{newtxtext}
\usepackage[varg]{newtxmath}

\usepackage{CJKutf8}
\usepackage{array}

\newcommand{\AmSLaTeX}{%
 $\mathcal A$\lower.4ex\hbox{$\!\mathcal M\!$}$\mathcal S$-\LaTeX}
\def\BibTeX{{\rmfamily B\kern-.05em
 \textsc{i\kern-.025em b}\kern-.08em
  T\kern-.1667em\lower.7ex\hbox{E}\kern-.125emX}}
\makeatletter
\def\tmpcite#1{\@ifundefined{b@#1}{\textbf{?}}{\csname b@#1\endcsname}}%
\makeatother
\field{D}
\vol{107}
\no{3}
\SpecialSection{Human Communication}
\title[Exploring the Effects of Japanese Font Designs on Impression Formation and Decision-Making in Text-Based Communication]
      {Exploring the Effects of Japanese Font Designs on Impression Formation and Decision-Making in Text-Based Communication}
\authorlist{%
 \authorentry[rintaro.chujo@hc.ic.i.u-tokyo.ac.jp]{Rintaro CHUJO}{n}{labelA}[labelA]\MembershipNumber{}
 \authorentry[atsuzuki@l.u-tokyo.ac.jp]{Atsunobu SUZUKI}{n}{labelA}[labelA]\MembershipNumber{}
 \authorentry[a.hautasaari@iii.u-tokyo.ac.jp]{Ari HAUTASAARI}{m}{labelA}[labelA]\MembershipNumber{2301008}
}

\affiliate[labelA]{The authors are with the University of Tokyo, Bunkyo-ku, 113-8654 Japan}
\received{2023}{5}{9}

\begin{document}
\begin{CJK*}{UTF8}{ipxm}
%\linenumbers
%%\linenumberdisplaymath
%%\realpagewiselinenumbers
%%\runninglinenumbers
%%\pagewiselinenumbers
\maketitle

\begin{summary}
Text-based communication, such as text chat, is commonly employed in various contexts, both professional and personal. However, it lacks the rich emotional cues present in verbal and visual forms of communication, such as facial expressions and tone of voice, making it more challenging to convey emotions and increasing the likelihood of misunderstandings. In this study, we focused on typefaces as emotional cues employed in text-based communication and investigated the influence of font design on impression formation and decision-making through two experiments. The results of the experiments revealed the relationship between Japanese typeface design and impression formation, and indicated that advice presented in a font evoking an impression of high confidence was more likely to be accepted than advice presented in a font evoking an impression of low confidence.
\end{summary}

\section{Introduction}
Nowadays, Computer-Mediated Communication (CMC) has become ubiquitous in various professional and personal contexts. Among CMC platforms, text-based chat applications, such as WhatsApp\footnote{https://www.whatsapp.com/} and WeChat\footnote{https://www.wechat.com/}, are particularly prevalent due to their simplicity in comparison to other forms of CMC, such as video conferencing systems.

Despite the prevalence of text-based communication via chat applications, there remain significant limitations in terms of the mediums' ability to convey emotions. The Media Richness Theory~\cite{daft_organizational_1986} posits that the availability of cues and properties varies depending on the method of communication, with text-based chat applications being particularly inadequate in facilitating the conveyance of emotional content. This is owing to the fact that text-based chats typically lack the rich emotional cues present in face-to-face interactions and video calls, as identified by Dennis \& Kinney~\cite{dennis_testing_1998}. Furthermore, the Social Information Processing Theory~\cite{walther_interpersonal_1992} suggests that misunderstandings are more likely to occur in situations where emotional cues are scarce, further exacerbating the problem of conveying emotional expressions in text-based communication.

In an effort to mitigate the limitations of text-based communication in conveying emotions, a number of solutions have been proposed to integrate paralinguistic emotional cues into chat applications. One emotional cue that has been the subject of study in recent years is typography. Studies have explored the correlation between font design and emotion, revealing that typography can convey the degree of emotional valence. However, much of this research has been focused on European typefaces~\cite{choi_emotype_2019,choi_typeface_2016,yonekura_2019}, and the relationship between Japanese typefaces, which exhibit more intricate shapes, has yet to be fully illuminated. Assessing Japanese typefaces is essential to determine the generalizability of the findings from conventional font research across cultures and languages.

Furthermore, typefaces hold the potential to not only act as emotional cues, but also to influence the decision-making of the message recipient, similarly to other emotional cues such as emoticons \cite{duan_emoticons_2018}. Specifically, although prior investigations have demonstrated the link between confidence and decision making~\cite{sniezek_cueing_1995,sniezek_trust_2001,van_swol_effects_2009,van_swol_factors_2005}, the precise relationship between the interpersonal impressions projected by typefaces and decision making has yet to be conclusively established in previous studies even with European typefaces. In everyday life, significant decisions are frequently made through communicative interactions rather than in isolation. With the increasing use of text-based chats, investigating the impact of paralinguistic emotional cues on decision-making is of paramount importance.

In this paper, we focus on the effects of Japanese font designs on both impression formation and decision-making. Firstly, we conducted a crowdsourced font evaluation experiment as Experiment 1 to examine the relationships between Japanese fonts and impressions of emotions and confidence levels. Secondly, based on the findings from Experiment 1, we identified the Japanese fonts that convey either high confidence or low confidence and performed Experiment 2 using the Judge-Advisor System~\cite{sniezek_cueing_1995} to examine the impact of font design on decision-making in text-based communication. The results of this experiment revealed that advice presented with a font that evokes an impression of high confidence was more likely to be accepted than advice presented with a font that evokes an impression of low confidence. Based on these findings, we discuss the role of fonts in text-based communication and propose future directions aimed at facilitating socio-emotional text-based communication and decision-making.

\section{Related Work}
\subsection{The Effects of Fonts on Emotional Communication in Text Chats}
There has been some prior research on the potential of typography to convey emotions, particularly in the context of European scripts. For instance, Choi, Yamazaki and Aizawa~\cite{choi_typeface_2016} demonstrated the capability of typography to convey emotions based on core affect~\cite{russell_core_2003} and basic emotions~\cite{ekman_are_1992}. Furthermore, previous research has also proposed text-based communication systems that leverage different typographical forms~\cite{choi_emotype_2019,yonekura_2019}, and it has been reported that altering the typographical form in text-based communication can be an effective means of expressing emotions~\cite{choi_typeface_2016}. 

In contrast, the majority of the empirical studies that have examined the relationship between font shape and emotion in CMC have primarily focused on fonts for the Latin alphabet, with limited examination of Japanese fonts. Many of the previous studies that have established a relationship between Japanese fonts and emotion have employed the semantic differential (SD) method~\cite{osgood_measurement_1957} to evaluate font impressions; however, the adjectives used in the evaluation were rather arbitrary, making it difficult to generalize the findings~\cite{Yang_2022}. To address these limitations, Toyosawa~\cite{toyosawa_2015} focused on core affect~\cite{russell_core_2003}, which comprises two dimensions of emotional valence and arousal that have been widely employed in traditional CMC research. However, one limitation of this examination was the inclusion of only five fonts.

In regards to the relationship between fonts and confidence, previous research has established a correlation in handwritten texts~\cite{warner_attributions_1986,shingaki_2009}. However, studies that have investigated the relationship between the design of fonts presented on computers and confidence are scarce.

As such, the primary objective of this study is to examine the relationship between Japanese fonts and impression formation by utilizing core affect and basic emotions as measures of emotion, and confidence as a measure of their influence on decision-making. The following research questions (RQs) were established:

\begin{quote}
RQ1: How consistent are the core affect and confidence impression ratings for Japanese fonts among raters?  
\end{quote}

\begin{quote}
RQ2: What type of Japanese fonts are associated with high and low impressions of  emotional valence (positive and negative),
arousal (aroused and calm), and confidence (confident and unconfident).  
\end{quote}

\begin{quote}
RQ3: What type of Japanese fonts are associated with each of the basic emotions?
\end{quote}

\subsection{Decision Making in Text Chats}
We conducted an examination of prior research on the utilization of paralinguistic emotional cues, such as emojis and typography, within text-based chat applications. Despite that numerous previous studies have substantiated that the incorporation of additional emotional cues enhances the conveyance of emotions, there remains a dearth of literature that goes into detail about the specific behavioral changes, especially decision-making, that result from the transmission of these emotions in text chats.

Much of the previous literature on the efficacy of online advice has been conducted within text-based communication that does not incorporate paralinguistic emotional cues (e.g., \cite{kock_psychobiological_2004,barak_comprehensive_2008,rummell_so_2010})
In comparison to other forms of text-based communication, such as e-mail, text-based chats possess the advantage of facilitating the conveyance of emotions through paralinguistic emotional cues such as emoticons and stamps, and there exists some evidence that impressions conveyed with these cues may influence decision-making. 
Prior research on the correlation between emotional cues and decision-making in text-based chat has primarily focused on the utilization of emoticons~\cite{duan_emoticons_2018}. However, it is important to note that emoticons, which often depict emotional facial expressions, possess limitations as emotional cues. That is, previous research has demonstrated that emoticons are commonly employed not only to express the sender's emotions but also to emphasize the content of the message.~\cite{thompsen_effects_1996,walther_interpersonal_1992,walther_impacts_2001}. Therefore, the relationship between the emotions conveyed in paralinguistic cues and the degree of acceptance of advice in text-based interactions remains uncertain.

On the other hand, there exists a large body of literature that has highlighted the impact of impressions regarding the sender of a message on decision-making. One measure that has been posited as a particularly influential factor is confidence. For instance, when the sender of a message conveys a numerical value of confidence in the advice in conjunction with the advice itself, message recipients are more likely to make a decision in accordance with the advice if they receive a higher confidence score~\cite{sniezek_cueing_1995,sniezek_trust_2001,van_swol_effects_2009,van_swol_factors_2005}. This is thought to be due to the propensity of humans to make decisions through the confidence heuristic~\cite{price_intuitive_2004}, whereby individuals rely on the perceived confidence level of a source when making decisions. It is plausible that when confidence is communicated through typography instead of a numerical value in text-based chats, the degree of advice acceptance may be affected in a similar manner as in previous studies. 

Hence, in this study, we focused on confidence as an impression about the other person that influences decision-making, and our second objective of this study was to investigate how decision-making is affected when confidence is communicated through typefaces within text-based chats. We hypothesized (H1): 

\begin{quote}
H1: Advice presented in a typeface that evokes a sense of confidence in the recipient is more likely to be accepted in comparison to advice presented in a typeface that evokes lack of confidence in the recipient.
\end{quote}

\section{Experiment 1: Evaluating the Design of Japanese Fonts as Emotional Cues}
Our fist experiment focuses on the utilization of typography as an emotional cue within text-based chat communication, with the aim of systematically examining the relationship between Japanese fonts, emotions, and confidence, which is known to impact decision-making, an area that has not been extensively explored.

\subsection{Task and Participants}
For the first experiment, stimuli were generated from the complete set of 93 fonts belonging to the Japanese font series "FONT1000 \footnote{http://www.font1000.com/}," included in the "Adobe Fonts \footnote{https://fonts.adobe.com/}" library. The complexity of the Japanese writing system and the abundance of characters present challenges in creating a diverse range of Japanese fonts. To circumvent these challenges, the FONT1000 series offers a wide variety of fonts by limiting the number of characters to 1,000 commonly used kanji, in addition to including all hiragana and katakana characters as well as punctuation marks. Additionally, the design of each typeface in the FONT1000 series ensures a consistent appearance of the characters, irrespective of their variations (kanji, hiragana or katakana).

\begin{figure}[t]
  \centering
  \includegraphics[width=\linewidth]{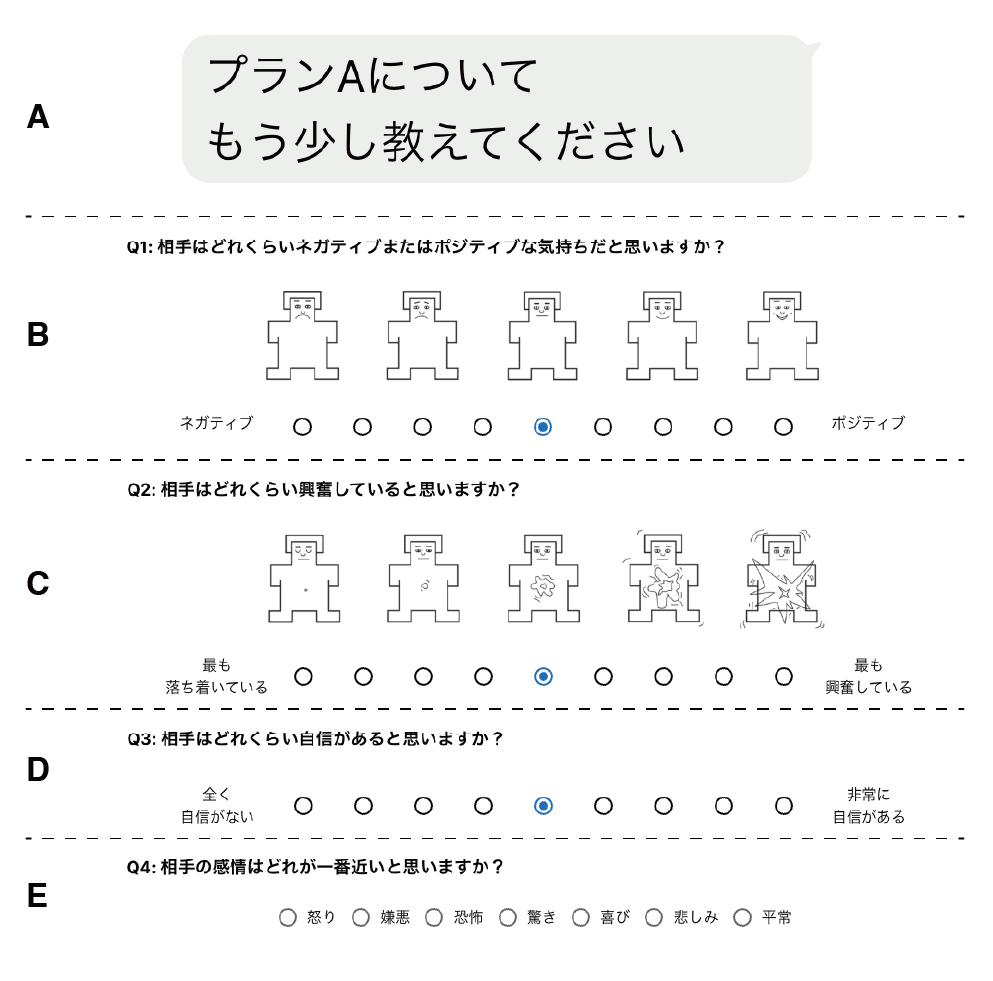}
  \caption{Schematic illustration of Experiment 1.}
  \label{fig:Schematic illustration of Experiment 1}
\end{figure}

During each trial of the experimental task, the message ``Please tell me more about Plan A" was presented in a Japanese font (Figure~\ref{fig:Schematic illustration of Experiment 1}A). The participants were instructed to consider this message as a reply to the question ``Which do you think is better, Plan A or Plan B?" They were then asked to rate their impressions of the sender of the message (Figure~\ref{fig:Schematic illustration of Experiment 1}A) in terms of core affect (Figure~\ref{fig:Schematic illustration of Experiment 1}B and Figure~\ref{fig:Schematic illustration of Experiment 1}C), confidence (Figure~\ref{fig:Schematic illustration of Experiment 1}D), and basic emotions (Figure~\ref{fig:Schematic illustration of Experiment 1}E). For the core affect evaluation, they were prompted to respond to the emotional valence and arousal level of the sender's emotions using the Self-Assessment Manikin (SAM) test \cite{bradley_measuring_1994}, which was used in a previous study on European fonts ~\cite{choi_typeface_2016}, on a 9-point scale (Figure~\ref{fig:Schematic illustration of Experiment 1}B and Figure~\ref{fig:Schematic illustration of Experiment 1}C), respectively. For the confidence assessment, the participants were prompted to respond to the level of confidence they perceived the sender felt using a 9-point scale ranging from ``not at all confident" to ``very confident" (Figure~\ref{fig:Schematic illustration of Experiment 1}D). In the basic emotion assessment, they were asked to select which category of the following: anger, disgust, fear, happiness, sadness, surprise, and neutral the sender's emotions fit the best (Figure~\ref{fig:Schematic illustration of Experiment 1}E). 

The experimental tasks were developed as web applications utilizing Next.js and Firebase Realtime Database, and the participants completed the tasks on a web browser via their own personal computers or other devices.

We hired 177 Japanese participants for this experiment (139 males and 36 females, $M_{age}=46.4$ years, $SD = 11.3$, excluding non-response) via a Japanese crowdsourcing platform Yahoo! Crowdsourcing \footnote{https://crowdsourcing.yahoo.co.jp/}. None of the participants had any background information about the experiment before joining the study. They were compensated for their participation in accordance with the rules of our university.

To assess the fonts, all 93 fonts were sorted into four sets, and the participants were tasked to evaluate each font once per set, while also responding to a concentration check test three times during the assessment of each set. The concentration check test was administered by displaying the message (Figure~\ref{fig:Schematic illustration of Experiment 1}A) as ``Please select the most positive emotion for the following question" in lieu of ``Please tell me more about Plan A", and the participants were instructed to respond using the 9-point Self-Assessment Manikin (SAM) scale.

Participants had the option to participate in the experiment for a minimum of one set and a maximum of five sets. Data from 167 participants were analyzed, as ten participants provided inaccurate responses in the concentration check test. 50 raters were recruited for each of the 93 fonts for a total of 4650 responses. Data from 10 participants who provided incorrect responses in the concentration check test were excluded. As a result, an average of 47.2 (45-49) participants per font, for a total of 4393 responses, were utilized in the analyses.

\begin{figure}[t]
  \centering
  \includegraphics[width=\linewidth]{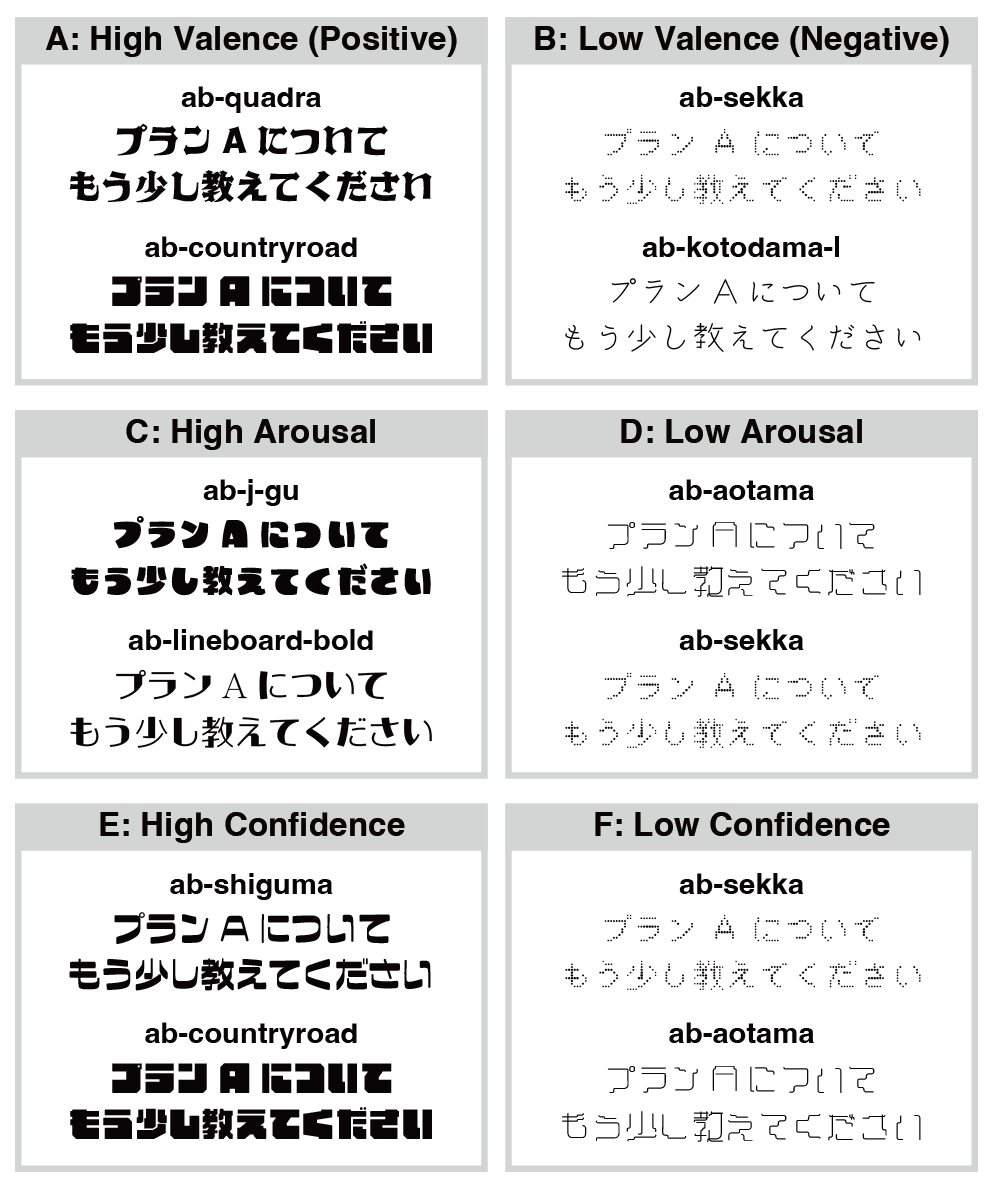}
  \caption{Examples of fonts with high and low impressions of emotional valence (positive and negative), arousal (aroused and calm), and confidence (confident and unconfident).}
  \label{fig:Examples of fonts with high and low average ratings on the emotional valence, arousal, and confidence axes.}
\end{figure}

\begin{figure}[t]
  \centering
  \includegraphics[width=\linewidth]{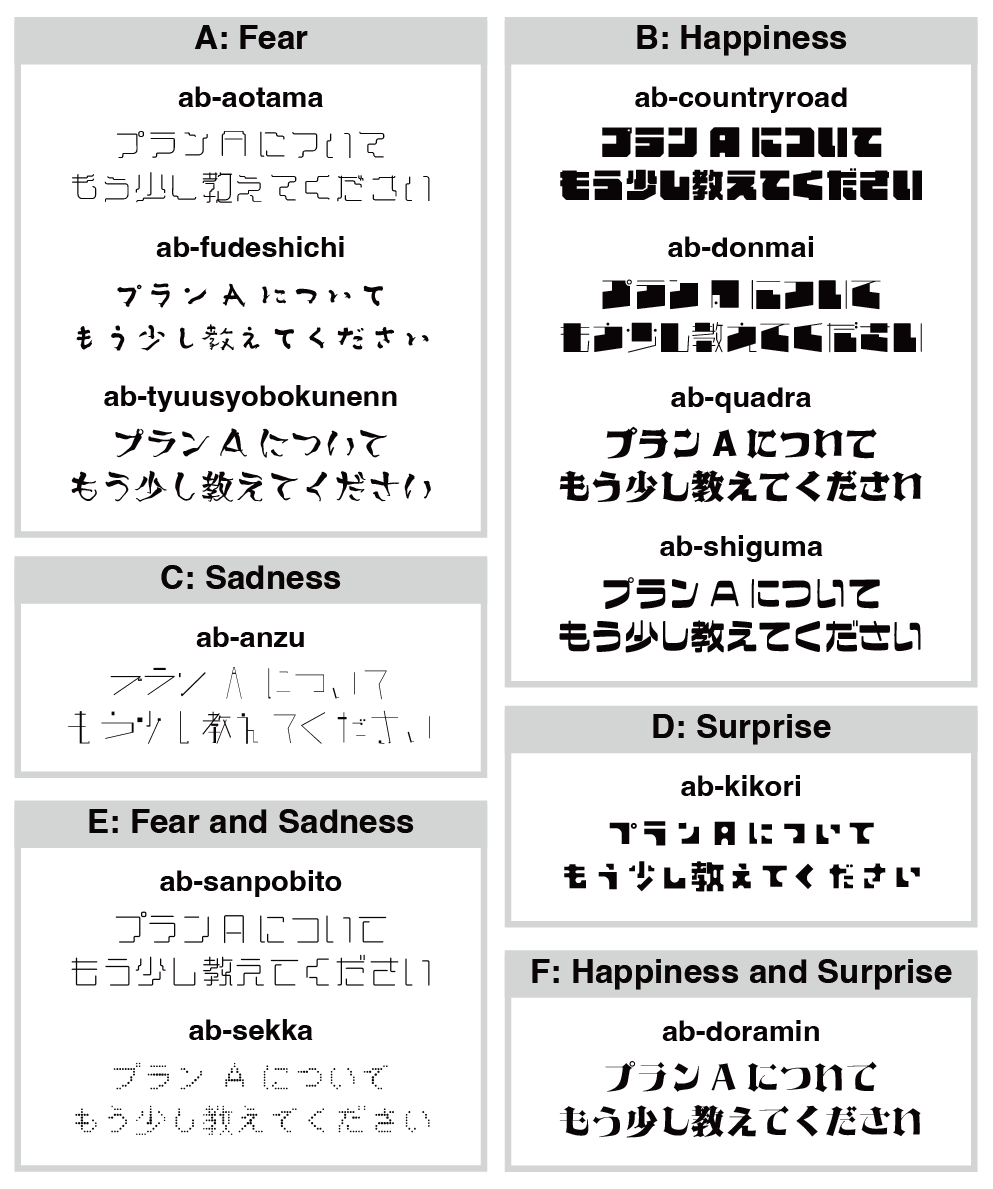}
  \caption{Examples of fonts where the selection rate for each basic emotion was significantly higher than chance level.}
  \label{fig:Examples of fonts where the selection rate for each basic emotion was significantly higher than the chance level}
\end{figure}

\subsection{Results}

Initially, to examine the consistency of impression ratings of emotion and confidence in the Japanese fonts across raters (RQ1), the correlation of each participant's impression ratings and the average of those of the other participants for each dimension of the core affect (emotional valence and arousal) and confidence were calculated following the process in previous work~\cite{martinez_quantifying_2020}. High inter-rater correlations were found for emotional valence and confidence, and moderate inter-rater correlation was found for arousal (emotional valence: $M = .568$, 95\% CI [.536, .600]; arousal: $M = .382$, 95\% CI [.339, .425]; confidence: $M = .553$, 95\% CI [.520, .588]).

Subsequently, in order to investigate which Japanese fonts were characteristic of each of the evaluation dimensions of emotional valence, arousal, and confidence (RQ2), the mean of each evaluation value for each font was calculated. Figure~\ref{fig:Examples of fonts with high and low average ratings on the emotional valence, arousal, and confidence axes.} illustrates two fonts with the highest and two fonts with the lowest mean assessment values for each dimension (valence, arousal and confidence).

In order to investigate which Japanese fonts were associated with each of the basic emotions (RQ3), we determined the selection rate of each basic emotion for each font and identified the fonts that had a significantly higher selection rate than chance level by a binomial test. Firstly, of the 93 fonts, we excluded those with significantly higher selection rates for "Neutral" compared to chance level, leaving 13 fonts. Among these, three fonts were associated with fear (Figure~\ref{fig:Examples of fonts where the selection rate for each basic emotion was significantly higher than the chance level}A), four with happiness (Figure~\ref{fig:Examples of fonts where the selection rate for each basic emotion was significantly higher than the chance level}B), and one each for sadness and surprise (Figure~\ref{fig:Examples of fonts where the selection rate for each basic emotion was significantly higher than the chance level}C and Figure~\ref{fig:Examples of fonts where the selection rate for each basic emotion was significantly higher than the chance level}D). Additionally, two fonts were associated with both fear and sadness (Figure~\ref{fig:Examples of fonts where the selection rate for each basic emotion was significantly higher than the chance level}E) and one font with both happiness and surprise (Figure~\ref{fig:Examples of fonts where the selection rate for each basic emotion was significantly higher than the chance level}F). We identified no fonts that were associated with anger or disgust at higher than chance level.

\subsection{Experiment 1: Discussion}
First, we examined the degree of consistency (RQ1) among raters in their impression ratings of emotional valence, arousal and confidence towards Japanese fonts by utilizing inter-rater correlation coefficients. The results revealed high inter-rater correlations for emotional valence and confidence. Notably, these correlation coefficients were comparable to those of facial impression evaluations~\cite{martinez_quantifying_2020}, which have been considered as having high inter-rater correlation. This finding is congruent with prior research on European fonts that demonstrated fonts to be effective in communicating emotional valence~\cite{choi_emotype_2019}, implying that fonts can serve as reliable emotional cues with consistent evalutions among raters.

Second, in order to investigate what type of Japanese fonts were associated with high and low impressions of emotional valence, arousal and confidence (RQ2), we analyzed the mean of each evaluation value for each font. Figure~\ref{fig:Examples of fonts with high and low average ratings on the emotional valence, arousal, and confidence axes.} illustrates that fonts that were rated positively in terms of emotional valence and high in confidence had thicker lines compared to fonts that were rated negatively and low in confidence. This finding is consistent with previous research~\cite{iwata_2003} which found that fonts with thin lines were associated with ``ambiguity". In future work, we are interested in investigating in detail the parameters (e.g., line thickness) that contribute to the emotional valence and confidence perceptions in Japanese fonts. Additionally, there was a tendency for fonts rated negatively in terms of emotional valence and low in confidence to have inconsistent line widths, as well as low readability. This is in contrast to previous research on European fonts~\cite{choi_typeface_2016} which found that complex and difficult-to-read fonts are not suitable for expressing emotion. Future research should investigate the relationship between readability and the emotional valence and confidence impressions of fonts.

Third, in order to investigate what type of Japanese fonts were associated with each basic emotion (RQ3), we determined the selection rate of each basic emotion for each font, and identified the fonts that had a significantly higher selection rate of basic emotions compared to chance level. The results revealed that fonts that only conveyed fear were characterized by non-uniformly thick lines and trembling lines (Figure~\ref{fig:Examples of fonts where the selection rate for each basic emotion was significantly higher than the chance level}A). This finding is consistent with previous research \cite{iwata_2003}, which found that fonts with wavy line thickness were associated with feelings of fear and worry. Conversely, fonts that conveyed both fear and sadness tended to have extremely thin lines, which is distinct from fonts that only conveyed fear (Figure~\ref{fig:Examples of fonts where the selection rate for each basic emotion was significantly higher than the chance level}E). This suggests that fonts may be capable of conveying complex emotions that are composed of Ekman's (1992) basic emotions. Additionally, all of the fonts that conveyed happiness were characterized by thicker lines, however, some of these fonts were difficult to read. Further investigation is required to explore this contradiction with previous research on European fonts~\cite{choi_typeface_2016}, which found that typefaces that are difficult to read are not suitable for expressing emotion. 

Lastly, it is important to highlight that only a small subset of the Japanese fonts in our dataset were associated with specific basic emotions. As such, our results indicate the ambiguity of standard font designs in terms of conveying basic emotions, specifically, but at the same time call to attention the opportunities for design of Japanese fonts to support emotional expression in text-chats, particularly in terms of negative emotions such as anger and disgust. 

\section{Experiment 2: Effect of Japanese Fonts on Decision Making in Text Chats}
\subsection{Task and Participants}

We conducted Experiment 2 to investigate the impact of font-communicated confidence on decision making. The experimental paradigm employed was the Judge-Advisor System~\cite{sniezek_cueing_1995}, a widely-utilized method in decision making research that allows for assessment of the degree of advice acceptance by comparing decisions made before and after receiving advice. 

The task used for decision making was the dot counting task~\cite{boldt_shared_2015,pescetelli_role_2021}. Participants were presented with a square shape on a computer screen containing an arbitrary number of dots on each side (randomly determined on each trial) of the gaze point. The number of dots on one side was controlled by the program, with $200 + d$ dots on either side and $200 - d$ dots on the other side. The dots were randomly distributed on a 20 x 20 invisible grid inside a square. The value of $d$ was manipulated by the 2-down-1-up method~\cite{levitt_transformed_1971} to ensure equal correct response rates for all participants. 

\begin{figure}[t]
  \centering
  \includegraphics[width=\linewidth]{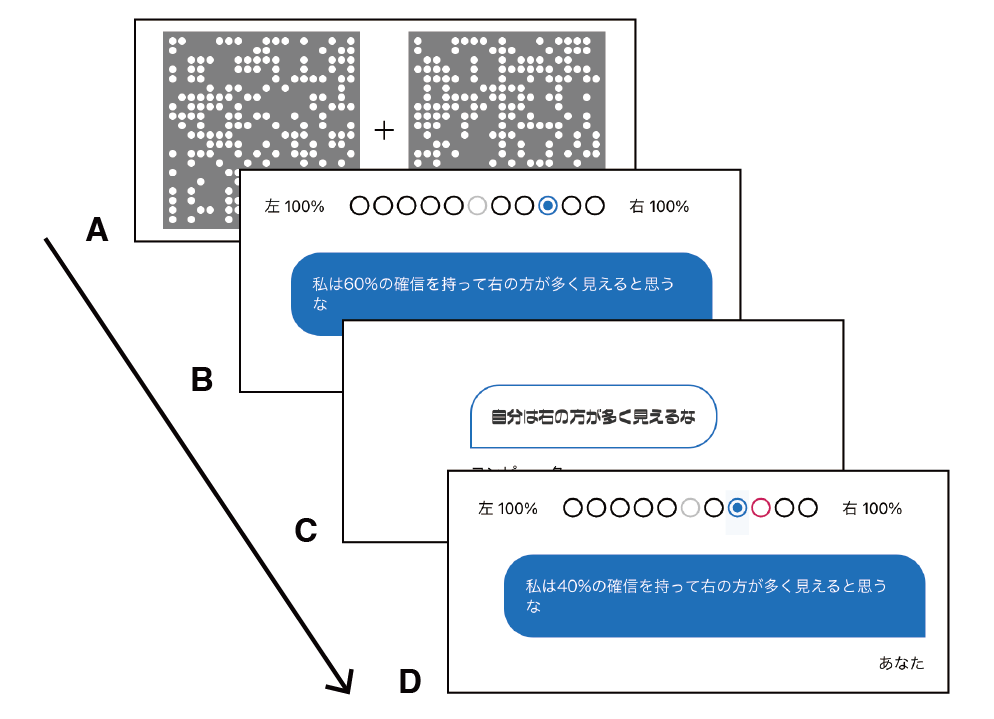}
  \caption{Schematic illustration of Experiment 2 paradigm.}
  \label{fig:Schematic illustration of Experiment 2 paradigm.}
\end{figure}

The flow of a single iteration of the dot counting task is depicted in Figure~\ref{fig:Schematic illustration of Experiment 2 paradigm.}. To initiate the task, participants were instructed to press the start button, after which the dot patterns were presented for a duration of 500 milliseconds on both sides of the screen (Figure~\ref{fig:Schematic illustration of Experiment 2 paradigm.}A). The participants were subsequently prompted to indicate which side they believed contained the greater number of dots with ten different options, ranging from ``I am 100\% sure it is left" to ``I am 100\% sure it is right", with increments of 20\% (Figure~\ref{fig:Schematic illustration of Experiment 2 paradigm.}B). A middle choice to indicate uncertain was not allowed. The participants were required to make their selection and indicate their level of confidence by clicking on the corresponding radio button with their mouse, and then clicking the ``Send" button.

Upon activation of the ``Send" button, the options were instantly obscured, and three seconds later, only the text message corresponding to the selected choice reappeared on the screen. Six seconds after the ``Send" button was activated, a statement in the form of ``I believe there are more dots on the right than on the left." or ``I believe there are more dots on the left than on the right." was displayed in Japanese, and in black letters on a white background, as purported advice from the computer (Figure~\ref{fig:Schematic illustration of Experiment 2 paradigm.}C). 

The advice was displayed in one of eight typefaces. The typefaces employed comprised of the four fonts that had the highest mean confidence scores (high confidence fonts) and the four fonts that had the lowest mean confidence scores (low confidence fonts) in Experiment 1. 

Finally, after reviewing the purported advice, the participants responded again using the same 10-point response box as before the advice (as depicted in Figure~\ref{fig:Schematic illustration of Experiment 2 paradigm.}D). At this time, the previously selected options were outlined in red to indicate the pre-advice response. Upon completion of the response, the next trial was immediately initiated.

The dot-counting task consisted of 48 trials, which were divided into six blocks. In each block, each of the eight fonts was used once in a randomized order as the font for the advice message. Thus, all fonts were presented six times during the experiment. Additionally, five practice trials were conducted prior to the initiation of the 48 primary trials. Prior to commencing the main trials and upon completion of the main trials, the participants were evaluated to ensure their concentration on the task. Specifically, during the dot pattern display, 400 dots were presented on one side and one dot on the other, and the participants were prompted to indicate which side they perceived to contain the greater number of dots. Similar assessments were conducted post the completion of the 48 main trials. 

The experimental tasks were developed as web applications utilizing Next.js and Firebase Realtime Database, and the participants completed the tasks on a web browser via their own personal computers or other devices.

We hired 100 Japanese paricipants for this experiment (80 males and 20 females, $M_{age}=46.3$ years, $SD = 11.0$) via the crowdsourcing platform Yahoo! Crowdsourcing \footnote{https://crowdsourcing.yahoo.co.jp/}. None of the participants had any background information about the experiment before joining the study. They were compensated for their participation in accordance with the rules of our university.

\subsection{Results}
Data analysis was performed utilizing data from 95 participants out of the total 100 after excluding two participants who did not fully complete the experiment and three participants whose responses remained unchanged prior and subsequent to receiving advice in all trials of the dot-counting task. Notably, no participants provided inaccurate responses during the concentration check task.

The acceptance of advice, denoted as $\delta_C$, was calculated in accordance with the methodology outlined in a previous study~\cite{pescetelli_role_2021} by utilizing the following formula for determining the difference in confidence between pre-advice and post-advice responses.

When the pre-advice answer and the advice direction were consistent:
\begin{equation}
\delta_C = C_{post} - C_{pre}
\end{equation}

When the answer before the advice and the direction of the advice were inconsistent:
\begin{equation}
\delta_C = C_{pre} - C_{post}
\end{equation}

$C_{pre}$ represents the confidence level of the response prior to receiving advice, ranging from 20\% to 100\%; i.e., $C_{pre}$ does not take into account whether participants thought the left or right side contained more dots. $C_{post}$ represents the confidence level of the response following advice, ranging from -100\% to 100\%. If the sides chosen by the participants changed between before and after the advice, $\delta_C$ was coded as negative, and conversely, as positive if the sides were unchanged. As a result, $\delta_C$ has a maximum value of 200\% (when a participant chooses either the left or right answer with 100\% confidence prior to receiving advice, receives advice that differs from their pre-advice answer and chooses a different orientation after receiving advice with 100\% confidence) and a minimum value of -200\% (when a participant chooses either the left or right answer with 100\% confidence, receives the same advice as their pre-advice answer and chooses a different orientation than before receiving advice with 100\% confidence).

Our hypothesis (H1) posited that the advice written in high confidence fonts would be more accepted than the advice written in low confidence fonts. To test H1, $\delta_c$ was subjected to a two-way ANOVA with font (high or low confidence) and direction of advice (whether the advice was consistent or inconsistent with the pre-advice response) as within-participant factors (Figure~\ref{fig:Relationship between the acceptance of advice, font type, and the direction of advice. Error bars indicate standard errors.}). Results revealed a significant main effect of font ($F[1, 94] = 7.90$, $p = .006$, $\eta^2_p = .078$), with the acceptance of advice response being greater for high confidence fonts than for low confidence fonts. H1 was supported. The main effect of direction of advice was also significant ($F[1, 94] = 53.72$, $p < .001$, $\eta^2_p = .364$), with the degree of advice acceptance being greater for advice that differed from the pre-advice response than for the same pre-advice response. The interaction effect between font type and degree of advice was not significant ($F[1, 94] = 0.63$, $p = .429$, $\eta^2_p = .007$).

\begin{figure}[ht]
  \centering
  \includegraphics[width=\linewidth]{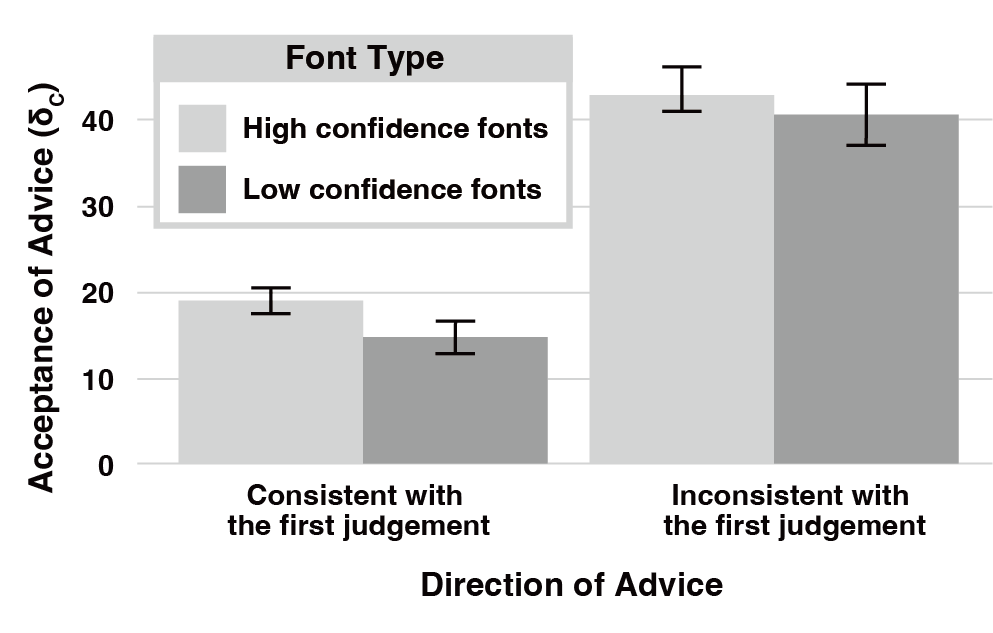}
  \caption{Relationship between the acceptance of advice, font type, and the direction of advice. Error bars indicate standard errors.}
  \label{fig:Relationship between the acceptance of advice, font type, and the direction of advice. Error bars indicate standard errors.}
\end{figure}

\subsection{Experiment 2: Discussion}

We examined the effect on the degree of advice acceptance when advice was given in text using Japanese fonts that conveyed the impression of high or low confidence (H1). 
We found that advice presented in a font that conveyed a high level of confidence was more likely to be accepted in comparison to advice presented in a font that conveyed a low level of confidence. In other words, this result suggested that the font used in text-based communication can influence the degree of advice acceptance by the recipient. This finding is congruent with previous research~\cite{sniezek_cueing_1995,sniezek_trust_2001,van_swol_effects_2009,van_swol_factors_2005}, which posited that when confidence is communicated numerically, advice is more likely to be accepted when a high level of confidence is conveyed. Our results imply that a ``confidence heuristic"~\cite{price_intuitive_2004} may be employed in decision-making even when fonts are employed to convey the level of confidence.

It is noteworthy that a significant main effect of advice direction was also observed in the experiment.
This result may can attributed to the experimental design. In this study, the degree of belief was set at five levels, ranging from 20\% to 100\% on the left and right sides, respectively, thus when the pre-advice response and the direction of the advice were the same, there was less scope for modifying the response in accordance with the advice than when the pre-advice response and the direction of the advice were dissimilar. 
Prior research on advice acceptance has established that decision makers tend to disregard advice due to a high level of confidence in their own judgment \cite{wang_why_2018}, and that the more divergent their own judgment and advice are, the less likely they are to accept the advice \cite{yaniv_receiving_2004}. Therefore, it is possible that advice that is congruent with the pre-advice response is more likely to be accepted than advice that is incongruent from the pre-advice response if an experimental design is employed that does not bias the maximum value of the degree of advice acceptance depending on the direction of the advice. Further research is required to determine the effect of advice direction on the degree of advice acceptance when the font is altered in text-based communication.

\section{Limitations and Future Work}

We conducted an evaluation experiment utilizing Japanese fonts and systematically examined the relationship between fonts, emotions and confidence. The results revealed a high inter-rater agreement in the evaluation of impressions of the fonts, which was established for the first time in this study using Japanese fonts. In future research, we intend to conduct similar experiments using European fonts to determine whether this result is specific to the Japanese language or not. Moreover, we also intend to explore imporessions about other paralinguistic emotional cues, such as speech balloons~\cite{aoki_emoballoon_2022}, which are known to convey emotional arousal in text-based communication. Additionally, considering the uneven gender balance in our studies due to the user demographics on the crowdsourcing platform~\cite{Nakagawa_yahoo_2014} and as also reported in previous research ~\cite{Sato_e_2022,Yamazaki_2020}, we are planning on incorporating gender and age as variables in future studies to further investigate any disparities in evaluation results depending on these characteristics.

Further research is also needed to ascertain whether the same results can be replicated for text-based communication in real-world settings, as opposed to a crowdsourcing study. With regard to the validity of the crowdsourcing experiments, previous research has demonstrated that effects can be replicated when crowdsourcing is employed in classical cognitive experiments, such as the Stroop effect~\cite{crump_evaluating_2013}. However, certain limitations have been identified, such as reduced statistical power when compared to laboratory experiments, as participants may not fully engage with the task~\cite{goodman_data_2013}. Therefore, we intend to first conduct a similar experiment in a laboratory setting to confirm the replicability of our results. In the future, we aim to confirm the robustness of the relationship between font shape and advice acceptance not only through experiments utilizing a computerized chatbot, but also through experiments that simulate a more natural environment and qualitative research methods such as focus groups.

\section{Conclusion}
In this study, we focused on Japanese fonts as emotional cues in text-based communication and confirmed the impact of font design on impression formation and decision-making. In Experiment 1, we conducted a font evaluation experiment to examine the relationship between Japanese fonts and core affect (emotional valence and arousal) and confidence, which is known to influence decision-making.  
In Experiment 2, we used the Judge-Advisor System~\cite{sniezek_cueing_1995} to examine the effect of font designs on the degree of advice acceptance. The results showed that advice presented in a font that conveyed a high level of confidence was more likely to be accepted than advice presented in a font that conveyed a low level of confidence. 

\section*{Acknowledgments}
This research is part of the results of Value Exchange Engineering, a joint research project between Mercari, Inc. and the RIISE. This work was supported by JSPS KAKENHI Grant Number JP18K18085. 

\bibliographystyle{ieicetr}
\bibliography{references}

\begin{thebibliography}{10}

\bibitem{daft_organizational_1986}
R.L. Daft and R.H. Lengel, ``Organizational {Information} {Requirements}, {Media} {Richness} and {Structural} {Design},'' Management Science, vol.32, no.5, pp.554--571, May\ 1986.

\bibitem{dennis_testing_1998}
A.R. Dennis and S.T. Kinney, ``Testing {Media} {Richness} {Theory} in the {New} {Media}: {The} {Effects} of {Cues}, {Feedback}, and {Task} {Equivocality},'' Information Systems Research, vol.9, no.3, pp.256--274, Sept.\ 1998.

\bibitem{walther_interpersonal_1992}
J.B. Walther, ``Interpersonal effects in computer-mediated interaction: A relational perspective,'' Communication Research, vol.19, no.1, pp.52--90, Feb.\ 1992.

\bibitem{choi_emotype_2019}
S.~Choi and K.~Aizawa, ``Emotype: {Expressing} emotions by changing typeface in mobile messenger texting,'' Multimedia Tools and Applications, vol.78, no.11, pp.14155--14172, June\ 2019.

\bibitem{choi_typeface_2016}
S.~Choi, T.~Yamasaki, and K.~Aizawa, ``Typeface {Emotion} {Analysis} for {Communication} on {Mobile} {Messengers},'' Proceedings of the 1st {International} {Workshop} on {Multimedia} {Alternate} {Realities}, {AltMM} '16, New York, NY, USA, pp.37--40, Association for Computing Machinery, Oct.\ 2016.

\bibitem{yonekura_2019}
R.~Yonekura, S.~Choi, R.~Yoshihashi, K.~Matsui, and A.~Hautasaari, ``Automated font selection system based on message sentiment in english text-based chat [in japanese],'' IEICE Technical Report, vol.118, no.502, pp.131--136, 2019.

\bibitem{duan_emoticons_2018}
J.~Duan, X.~Xia, and L.M. Van~Swol, ``Emoticons' influence on advice taking,'' Computers in Human Behavior, vol.79, pp.53--58, Feb.\ 2018.

\bibitem{sniezek_cueing_1995}
J.A. Sniezek and T.~Buckley, ``Cueing and {Cognitive} {Conflict} in {Judge}-{Advisor} {Decision} {Making},'' Organizational Behavior and Human Decision Processes, vol.62, no.2, pp.159--174, May\ 1995.

\bibitem{sniezek_trust_2001}
J.A. Sniezek and L.M. Van~Swol, ``Trust, confidence, and expertise in a judge–advisor system,'' Organizational Behavior and Human Decision Processes, vol.84, no.2, pp.288--307, March\ 2001.

\bibitem{van_swol_effects_2009}
L.M. Van~Swol, ``The {Effects} of {Confidence} and {Advisor} {Motives} on {Advice} {Utilization},'' Communication Research, vol.36, no.6, pp.857--873, Dec.\ 2009.

\bibitem{van_swol_factors_2005}
L.M. Van~Swol and J.A. Sniezek, ``Factors affecting the acceptance of expert advice,'' British Journal of Social Psychology, vol.44, no.3, pp.443--461, Sept.\ 2005.

\bibitem{russell_core_2003}
J.A. Russell, ``Core affect and the psychological construction of emotion,'' Psychological Review, vol.110, no.1, pp.145--172, Jan.\ 2003.

\bibitem{ekman_are_1992}
P.~Ekman, ``Are there basic emotions?,'' Psychological Review, vol.99, no.3, pp.550--553, July\ 1992.

\bibitem{osgood_measurement_1957}
C.E. Osgood, G.J. Suci, and P.H. Tannenbaum, The {Measurement} of {Meaning}, University of Illinois Press, Champaign, IL, USA, 1957.

\bibitem{Yang_2022}
N.~Yang, ``フォントの印象評価方法に関する考察 [in japanese],'' 日本デザイン学会研究発表大会概要集, vol.69, pp.32--33, 2022.

\bibitem{toyosawa_2015}
S.~Toyosawa, K.~Takashi, and M.~Hiroyuki, ``フォントデザインが読み手に与える情動への影響 [in japanese],'' 人間工学, vol.51, no.Supplement, pp.S320--S321, 2015.

\bibitem{warner_attributions_1986}
R.M. Warner and D.B. Sugarman, ``Attributions of personality based on physical appearance, speech, and handwriting,'' Journal of Personality and Social Psychology, vol.50, no.4, pp.792--799, April\ 1986.

\bibitem{shingaki_2009}
S.~Noriko and T.~Yukie, ``人は手書き文字をどのような次元で認知しているのか? [in japanese],'' 成城大学社会イノベーション研究, vol.4, no.2, pp.27--43, March\ 2009.

\bibitem{kock_psychobiological_2004}
N.~Kock, ``The {Psychobiological} {Model}: {Towards} a {New} {Theory} of {Computer}-{Mediated} {Communication} {Based} on {Darwinian} {Evolution},'' Organization Science, vol.15, pp.327--348, 2004.
\newblock Place: US Publisher: Institute for Operations Research \& the Management Sciences (INFORMS).

\bibitem{barak_comprehensive_2008}
A.~Barak, L.~Hen, M.~Boniel-Nissim, and N.~Shapira, ``A {Comprehensive} {Review} and a {Meta}-{Analysis} of the {Effectiveness} of {Internet}-{Based} {Psychotherapeutic} {Interventions},'' Journal of Technology in Human Services, vol.26, no.2-4, pp.109--160, July\ 2008.
\newblock Publisher: Routledge.

\bibitem{rummell_so_2010}
C.M. Rummell and N.R. Joyce, ``“{So} wat do u want to wrk on 2day?”: {The} ethical implications of online counseling,'' Ethics \& Behavior, vol.20, pp.482--496, 2010.
\newblock Place: United Kingdom Publisher: Taylor \& Francis.

\bibitem{thompsen_effects_1996}
P.A. Thompsen and D.A. Foulger, ``Effects of pictographs and quoting on flaming in electronic mail,'' Computers in Human Behavior, vol.12, no.2, pp.225--243, June\ 1996.

\bibitem{walther_impacts_2001}
J.B. Walther and K.P. D'Addario, ``The impacts of emoticons on message interpretation in computer-mediated communication,'' Social Science Computer Review, vol.19, no.3, pp.324--347, Aug.\ 2001.

\bibitem{price_intuitive_2004}
P.C. Price and E.R. Stone, ``Intuitive evaluation of likelihood judgment producers: evidence for a confidence heuristic,'' Journal of Behavioral Decision Making, vol.17, no.1, pp.39--57, Jan.\ 2004.

\bibitem{bradley_measuring_1994}
M.M. Bradley and P.J. Lang, ``Measuring emotion: {The} self-assessment manikin and the semantic differential,'' Journal of Behavior Therapy and Experimental Psychiatry, vol.25, no.1, pp.49--59, March\ 1994.

\bibitem{martinez_quantifying_2020}
J.E. Martinez, F.~Funk, and A.~Todorov, ``Quantifying idiosyncratic and shared contributions to judgment,'' Behavior Research Methods, vol.52, no.4, pp.1428--1444, Aug.\ 2020.

\bibitem{iwata_2003}
Y.~Iwata, M.~Iwata, and S.~Tano, ``フォント形状・感情・感性の相互依存関係の分析と関連ルール抽出の試み [in japanese],'' 感性工学研究論文集, vol.3, no.1, pp.7--16, 2003.

\bibitem{boldt_shared_2015}
A.~Boldt and N.~Yeung, ``Shared neural markers of decision confidence and error detection,'' The Journal of Neuroscience: The Official Journal of the Society for Neuroscience, vol.35, no.8, pp.3478--3484, Feb.\ 2015.

\bibitem{pescetelli_role_2021}
N.~Pescetelli and N.~Yeung, ``The role of decision confidence in advice-taking and trust formation,'' Journal of Experimental Psychology. General, vol.150, no.3, pp.507--526, March\ 2021.

\bibitem{levitt_transformed_1971}
H.~Levitt, ``Transformed {Up}‐{Down} {Methods} in {Psychoacoustics},'' The Journal of the Acoustical Society of America, vol.49, no.2B, pp.467--477, Feb.\ 1971.

\bibitem{wang_why_2018}
X.~Wang and X.~Du, ``Why {Does} {Advice} {Discounting} {Occur}? {The} {Combined} {Roles} of {Confidence} and {Trust},'' Frontiers in Psychology, vol.9, p.2381, 2018.

\bibitem{yaniv_receiving_2004}
I.~Yaniv, ``Receiving other people's advice: {Influence} and benefit,'' Organizational Behavior and Human Decision Processes, vol.93, no.1, pp.1--13, Jan.\ 2004.

\bibitem{aoki_emoballoon_2022}
T.~Aoki, R.~Chujo, K.~Matsui, S.~Choi, and A.~Hautasaari, ``Emoballoon - conveying emotional arousal in text chats with speech balloons,'' Proceedings of the 2022 CHI Conference on Human Factors in Computing Systems, CHI '22, New York, NY, USA, Association for Computing Machinery, 2022.

\bibitem{Nakagawa_yahoo_2014}
M.~Nakagawa, ``{マイクロタスク特化型Yahoo}!クラウドソーシングの現在と今後の展望,'' 赤門マネジメント・レビュー, vol.13, no.7, pp.263--274, 2014.

\bibitem{Sato_e_2022}
H.~Sato and K.~Omine, ``eスポーツの印象に影響を及ぼす要因の検討,'' 日本教育心理学会総会発表論文集, vol.64, p.274, 2022.

\bibitem{Yamazaki_2020}
I.~Yamazaki, R.~Ito, K.~Hamano, S.~Nakamura, A.~Kake, and K.~Ishimaru, ``記憶対象の文字の太さの違いが記憶容易性に及ぼす影響,'' 研究報告ユビキタスコンピューティングシステム（UBI）, vol.2020-UBI-68, no.22, pp.1--8, Dec.\ 2020.

\bibitem{crump_evaluating_2013}
M.J.C. Crump, J.V. McDonnell, and T.M. Gureckis, ``Evaluating {Amazon}'s {Mechanical} {Turk} as a {Tool} for {Experimental} {Behavioral} {Research},'' PLOS ONE, vol.8, no.3, p.e57410, March\ 2013.

\bibitem{goodman_data_2013}
J.K. Goodman, C.E. Cryder, and A.~Cheema, ``Data {Collection} in a {Flat} {World}: {The} {Strengths} and {Weaknesses} of {Mechanical} {Turk} {Samples},'' Journal of Behavioral Decision Making, vol.26, no.3, pp.213--224, April\ 2013.

\end{thebibliography}

\profile[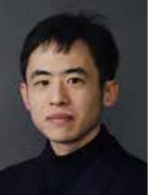]{Rintaro Chujo}{is currently a Master's student at the Graduate School of Interdisciplinary Information Studies, the University of Tokyo. He received his B.A. degree in Psychology from the University of Tokyo in 2023. His reserach interests are in human-computer interaction, affective computing and computer-mediated communication}
\label{profile}

\profile[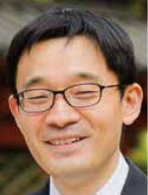]{Atsunobu Suzuki}{is currently an Associate Professor at the Department of Psychology, Graduate School of Humanities and Sociology, the University of Tokyo. He majored in Cognitive and Behavioral Sciences at the University of Tokyo, where he received his M.Sc. and Ph.D. His main research interests are person perception and aging.}

\profile[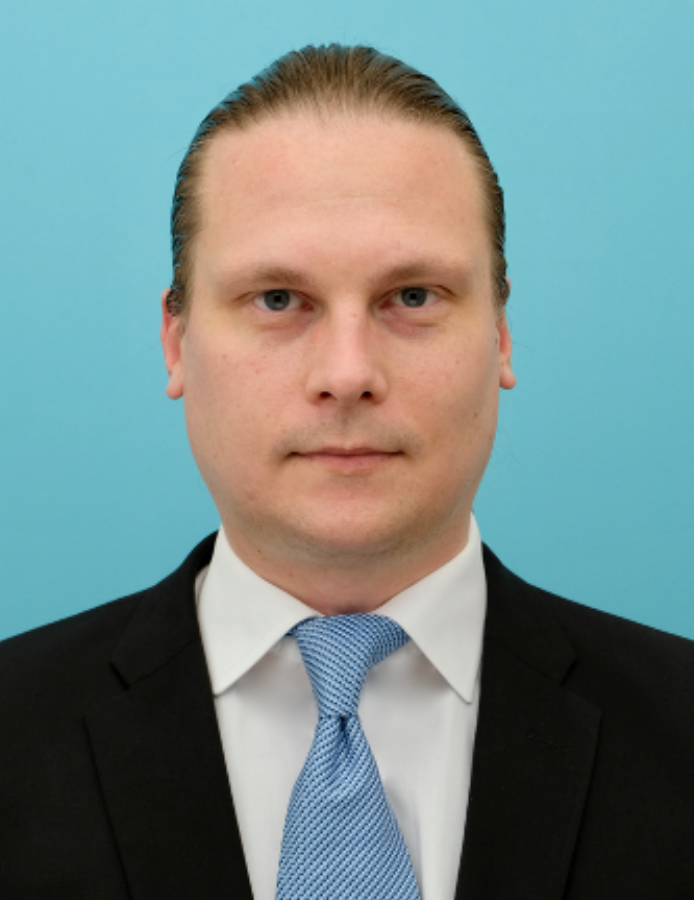]{Ari Hautasaari}{is currently a Project Associate Professor (RIISE - Research Institute for an Inclusive Society through Engineering) in the Interfaculty Initiative in Information Studies at the University of Tokyo. He received his M.Sc. in Information Processing Science from the University of Oulu, Finland, and his Ph.D. in Informatics from Kyoto University, Japan. His research interests include human-computer interaction, computer-mediated communication, and socio-emotional communication support.}

\end{CJK*}

\end{document}